\documentclass[aps,prl,twocolumn,showpacs,amsmath,amssymb]{revtex4}
\usepackage{graphicx}
\usepackage{dcolumn}
\usepackage{bm}

\begin{document}

\title{Scaling of excitons in carbon nanotubes}

\author{Vasili Perebeinos, J. Tersoff, and Phaedon Avouris$^*$}
\affiliation{IBM Research Division, T. J. Watson Research Center,
Yorktown Heights, New York 10598}

\date{\today}

\begin{abstract}
Light emission from carbon nanotubes is
expected to be dominated by excitonic recombination. Here we
calculate the
properties of excitons in nanotubes embedded in a dielectric, for
a wide range of tube radii and dielectric environments.  We find
that simple scaling relationships give a good description of the
binding energy, exciton size, and oscillator strength.
\end{abstract}

\maketitle

The optical properties of carbon nanotubes have
received increasing experimental and theoretical attention.
Optical absorption and emission spectra of carbon nanotubes have
been studied by a number of groups
\cite{Li,Connell,Bachilo,Hagen,Lebedkin,Lefebvre}; and electro-optical
devices have already appeared \cite{Misewich,Marcus}. Initial
attempts to explain the experimental observations naturally took
independent-electron theory as their starting point. However,
theoretically it is now clear that emission is dominated by
excitonic recombination \cite{Ando,Pedersen,Kane,Louie}.

A number of theoretical approaches have been used to describe
these excitons. One approach involves variational calculations
\cite{Pedersen,Kane}.  While valuable, these have been limited to
an effective-mass approximation,
and do not address issues of spectral weight.
The most accurate description is provided by an {\it
ab initio} solution of the Bethe-Salpeter equation using
GW-corrected quasi-particle energies \cite{Louie}.
However, it is not currently feasible to apply this
computationally intensive approach to a wide range of nanotube
sizes or environments.

Here we use an intermediate level of theory to
provide a broad overview of the exciton properties.
We calculate the excitonic properties of nanotubes
embedded in dielectric media, for the range of tube radii
and dielectric constants most relevant to potential applications.
We find that the exciton size,
binding energy, and oscillator strength all exhibit robust
(though approximate) scaling relationships.
The relationships obtained for the excitonic properties can be
used to better understand and optimize the operation of nanotube
opto-electronic devices.

The proper procedure for the calculation of excitons has been
described in detail in Ref.~\cite{Rohlfing}. It involves solving
the Bethe-Salpeter equation,
\begin{eqnarray}
\Delta_{k}A_{k}^S+\sum_{k'}{\cal K}_{k,k'}A_{k'}^S=\Omega_S
A_{k}^S \label{eq34}
\end{eqnarray}
where the kernel ${\cal K}_{k,k'}$ describes the interaction
between all possible electron-hole pairs of total momentum
$q_{exc}$, and $\Delta_k$ is the quasiparticle energy for a
non-interacting electron and hole with wavevector $k$ and
$q_{exc}-k$. The exciton momentum $q_{exc}$ is equal to that of
the exciting photon, and is hereafter approximated by $q_{exc}=0$.
We approximate the quasiparticle energies by eigenvalues of the
tight-binding Hamiltonian \cite{Saito,Louie2} ($t=3.0$ eV), with any
additional self-energy corrections restricted to the so-called
``scissors operator'', in which the self-energy is approximated by
a rigid shift of the conduction band relative to the valence band.
Since the quasi-particle bandstructures are not well known for
nanotubes of varying diameters and embedding media, we report only
properties that are not affected by the magnitude of this shift.

For the optically active singlet excitons, the
interaction has two contributions, direct (${\cal K}^d$)
and exchange (${\cal K}^x$):
\begin{eqnarray}
{\cal K}_{k,k'}&=&{\cal K}_{k,k'}^d+2{\cal K}^x_{k,k'}
\label{eq35}
\end{eqnarray}
where the direct (exchange) term is evaluated with the screened
(bare) Coulomb interactions \cite{Rohlfing}.
The unscreened Coulomb interaction between carbon p$_z$ orbitals
is modelled by the Ohno
potential, which realistically describes organic polymer systems:
\begin{eqnarray}
V(r_{ij})=\frac{U}{\sqrt{
\left(\frac{4\pi\varepsilon_0}{e^2}Ur_{ij}\right)^2+1}}
\end{eqnarray}
where $r_{ij}$ is the distance between sites $i$ and $j$, and
$U=11.3$ eV is the energy cost to place two electrons on a single
site ($r_{ij}=0$). Our results are not sensitive to
the value of $U$ when the size of the exciton is large.

An ideal calculation would include the nonlocal dielectric
response of both the nanotube itself and the medium in which it is
embedded.  Here, for computational simplicity, we replace this
complicated response function with a single dielectric constant
$\varepsilon$ \cite{dielectric}. This is most accurate for narrow
tubes, and for embedding media with large dielectric constants.
In this regime, the exciton length along the tube is large
relative to the tube radius, and most of the dielectric screening
occurs in the surrounding medium. The screening is then well
described by the dielectric constant $\varepsilon$ of the nanotube
environment. For isolated nanotubes, or tubes in low-$\varepsilon$
media, this treatment is not very accurate. Fortunately, it
is most accurate in precisely the regime of greatest technological
interest. Modulated electro-optical devices \cite{Misewich,Marcus}
are most practical for relatively narrow tubes embedded in SiO$_2$
($\varepsilon \sim 4$) or higher-$\varepsilon$ materials.

To calculate the optical properties, we evaluate the imaginary part of the
dielectric function for light polarized along the nanotube axis
\cite{DelSole}:
\begin{eqnarray}
\epsilon_2(\omega)&=&\frac{8\pi^2
e^2}{V_0m_e^2}\sum_S\left|\sum_{k}A_{k}^S\frac{P_{cv}(k)}
{\Delta_k}\right|^2 \delta(\hbar\omega-\Omega_S) \label{eps2}
\end{eqnarray}
where $P$ is the dipole matrix element
\cite{dipole}. The optical response Eq.~(\ref{eps2}) is the same
as derived in the presence of the $GW$ non-local potential
\cite{DelSole}.
$\varepsilon_2(\omega)$ obeys  a sum
rule, where $\int\varepsilon_2d\omega\propto\sum_k
P_{cv}^2(k)/\Delta^2_k$ is a constant independent of the strength
of the screened interaction $e^2/\varepsilon$.

\begin{figure}
\includegraphics[height=2.34in,width=2.85in,angle=0]{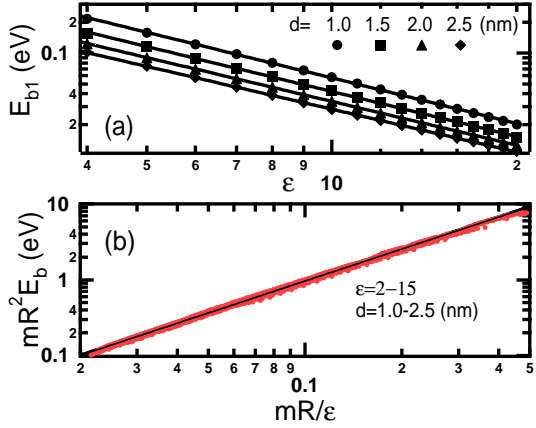}
\caption{\label{fig1} (a) Binding energy of first optically
active exciton, vs.\ $\varepsilon$,
in four semiconducting zig-zag tubes: (13,0),
(19,0), (25,0), and (31,0). (b) Scaling of binding energy of
first and second exciton (red dots) in
semiconducting tubes with all possible chirality  (156 tubes with
$d=1.0-2.5$ nm, $\varepsilon=2-15$).
Here $R$ and $m$ are in a.u.
The black solid  line is the
best fit to Eq.~(\protect{\ref{eq47}}) for $\varepsilon=4-15$
(3\% RMS error over this range),
corresponding to $\alpha=1.40$ and ${\rm A_b}=24.1$ eV.}
\end{figure}

We solve the BSE equation (\ref{eq34}) by direct diagonalization,
choosing a $k$ sample sufficient to converge the low-energy
optical spectra (and {\it a fortiori} the binding energies).
We calculate the binding energy, size, and spectral function for
singlet excitons.
The binding energy of the first optically active exciton
vs.\ $\varepsilon$ is shown on Fig.~\ref{fig1}a for four
zig-zag tubes with diameters $d=1.0-2.5$ nm.
(There is another singlet state 3-5 meV lower in energy,
but it is optically silent by symmetry.)

The dependence of binding energy $E_{b1}$ on $\varepsilon$ in
Fig.~\ref{fig1} can be fitted well with a power law, with the
exponent being almost independent of the tube diameter.  This
suggests a more general power law scaling, which we can motivate
in an effective mass approximation as follows.
Given a variation
wavefunction described by a single parameter $L$ that scales the
size along the tube axis, the exciton binding energy is:
\begin{eqnarray}
E_{L}&\propto&\frac{\hbar^2}{2mL^2}-\frac{e^2}{\epsilon
R}f\left(\frac{L}{R}\right)=\frac{\hbar^2}{mR^2}
g\left(\frac{L}{R},\frac{mR}{\varepsilon}\right) \label{eq46}
\end{eqnarray}
Here the first term is the kinetic energy,
and $m$ is the effective mass.
The second term is the potential energy,
which depends on the exciton size {\it via} the
dimensionless function $f(L/R)$.
Then the exciton binding energy is
\begin{eqnarray}
E_{b}&=&\min_{L}\left( E_{L} \right)
=\frac{\hbar^2}{mR^2} h\left(\frac{mR}{\varepsilon}\right)
\label{eq47a}\\
E_{b}&\approx& {\rm A_b} R^{\alpha-2}
m^{\alpha-1}\epsilon^{-\alpha}
\label{eq47}
\end{eqnarray}
where we approximate the function $h$ by a power law over the
range of interest, with empirical parameters $\alpha$ and $A_b$.
The effective mass $m$ depends on the tube indices \cite{Saito}
(i.e.\ on radius and chirality).

In 3D semiconductors, the potential energy is $\propto 1/L$, and
the energy is minimized when the exciton size
$L_S\propto\varepsilon / m$,
so the binding energy scales as $E_b\propto m/\epsilon^2$.
This corresponds to true scaling, with a power law $\alpha=2$ in
Eq.~(\ref{eq47}). In the case of nanotubes, the power-law scaling
is only an approximation. Nevertheless, for the most important
range of tube sizes and dielectric constants, the behavior is
rather well described by a power-law scaling in $R$ and
$\varepsilon$ with a single value of $\alpha$.
Indeed, all the binding energies for the first and second excitons
in semiconducting tubes, with all possible chiralities,
($d=1.0-1.5$ nm) collapse onto a single curve shown in
Fig.~\ref{fig1}b. Similar energy scaling was reported by Pedersen
\cite{Pedersen} in a variational effective-mass model.
In the range $\varepsilon\gtrsim 4$, where our approach is most reliable,
we obtain the best fit with
$\alpha=1.40$.

The second exciton that is optically active derives
primarily from the second band of the nanotube. It falls within
the continuum of the first band, and so becomes a resonance with a
finite lifetime \cite{Louie}. By artificially turning off the
interband coupling, we determine that this coupling has very
little effect on the exciton energy.

Considerable attention has been focused on the
ratio $E_{b2}/E_{b1}$ between the binding energies
of the first and second excitons \cite{Kane,Louie}.
($E_{b2}$ is defined relative to the second-band
quasiparticle gap.  Note that the exciton {\it formation} energies
involve also the {\it quasiparticle} bandgaps.)
The scaling relation
of Eq.~(\ref{eq47}) predicts
$E_{b2}/E_{b1}=(m_2/m_1)^{\alpha-1}$, where $m_2$ and $m_1$ are
the effective masses of the first and second bands. In the case of
zig-zag tubes:
\begin{eqnarray}
m_1&=&\frac{\hbar^2\Delta_1}{3a^2t^2}
\left(1+\sigma\frac{\Delta_1}{2t}\right)^{-1}
\nonumber \\
m_2&=&\frac{\hbar^2\Delta_2}{3a^2t^2}
\left(1-\sigma\frac{\Delta_2}{2t}\right)^{-1}
\label{mass}
\end{eqnarray}
Here $\Delta_1$ and $\Delta_2$ are the tight-binding bandgaps;
$a$ is the graphene lattice constant;
and for tube indeces (n,0),
$\sigma$=1 if mod(n,3)=1 and $\sigma$=-1 if mod(n,3)=2.

It is common to treat the gap values in the infinite-radius limit,
$\Delta_2^{\infty}=2\Delta_1^{\infty}=2ta/\sqrt{3}R$.
This is rather accurate (within 5\%) for
tubes with $d=1.0-2.5$. On the other hand,
for the same range of diameters, the effective mass ratio $m_2/m_1$
varies from 3.4 to 1.3, approaching the infinite-radius limit
$m_2/m_1\rightarrow 2$ much more slowly than the gap ratio.
Thus caution must be used in discussing available experimental data
in terms of the $R\rightarrow\infty$ limit \cite{Kane}.
In
particular, for the (8,0) tube $m_2/m_1=0.96$, and according to
Eq.~(\ref{eq47}) the binding energies of the first two excitons
should be very similar. Indeed, the accurate first-principles
calculations by Spataru et. al. \cite{Louie} find the binding
energies of the first and second excitons (A'$_1$ and C'$_1$ in
\cite{Louie}) to be 0.99 and 1.00 eV. Using $\varepsilon=1.93$ to
best reproduce this, our calculations give $E_{b1}=0.99$ eV and
$E_{b2}=1.05$ eV. In contrast, the (10,0) tube has $m_2/m_1=4.14$,
and for the same $\varepsilon=1.93$ we find $E_{b2}/E_{b1}=1.41$.
[The simple $m^{\alpha-1}$ scaling is not accurate for such small
$\varepsilon$.]
It is important to note that effective mass
dependence similar to Eq.~(\ref{mass}) holds also for chiralities
other from zig-zag tubes.  Thus we expect exciton properties
in tubes of index (m,n) to depend primarily on whether
mod(n-m,3)=1 or 2, independent of the chiral angle.

\begin{figure}
\includegraphics[height=2.34in,width=2.85in,angle=0]{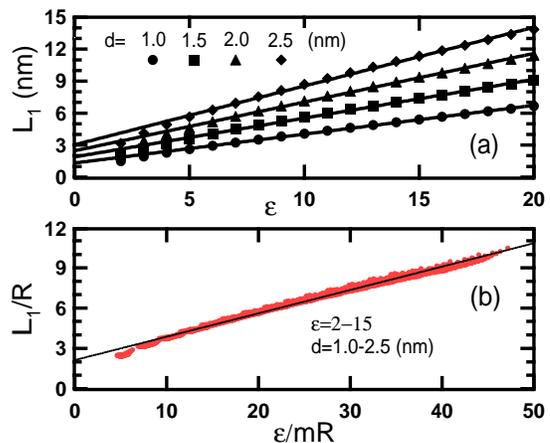}
\caption{\label{fig2} (a) First exciton RMS e-h separation $L_1$ in
four zig-zag tubes (13,0), (19,0), (25,0), and (31,0). The solid
lines are the best linear fits.
The slope $k$ scales approximately as $m^{-1}$:
the product $km_1$ (in a.u.) equals 0.167, 0.160, 0.157,
and 0.155 for tubes  with diameter 1.0, 1.5, 2.0, and
2.5 nm, respectively.
(b) The linear scaling of $L_1/R$ with
$\varepsilon/mR$ in semiconducting tubes
of all possible chirality with $d=1.0-2.5$ nm,
and $\varepsilon=2-15$. The solid line
is the best fit to Eq.~(\protect{\ref{LL1}}) to the
results in the range $\varepsilon\ge 4$,
giving $A_L=2.13$ and $B_L=0.174$ (for $m$ and $R$ in a.u.).
The RMS discrepancy between the
fit and the full calculations is 2\% for the subset of data
having $\varepsilon\geq 4$ or $\varepsilon/mR^2\geq 9.5$.}
\end{figure}

To quantify the exciton size, we use the root-mean-square (RMS)
distance between electron and hole, $L_S$. The size $L_1$ of the
first exciton is shown in Fig.~\ref{fig2}a as a function of
$\varepsilon$, for four different tube diameters. The size is
approximately linear in $\varepsilon$. From Eq.~(\ref{eq46}), $L_S
/ R$ is expected to be a function of $mR/\varepsilon$. Combining
this with the observed linear dependence on $\varepsilon$, we
anticipate that the exciton size will obey the scaling
relationship:
\begin{eqnarray}
\frac{L_1}{R}={\rm A_L}+{\rm B_L}\frac{\varepsilon}{Rm_{1}}
\label{LL1}
\end{eqnarray}
This is confirmed in Fig.~\ref{fig2}b, which shows a linear
dependence of $L_1/R$ on $\varepsilon/m_{1}R$ in all
semiconducting tubes with $d=1.0-2.5$ nm, for all chiralities.

The exciton size directly affects observable quantities such as
the exciton oscillator strength and the radiative lifetime. The
exciton oscillator strength is proportional to the probability to
find an electron and a hole at the same position \cite{Elliot}. In
3D semiconductors this is inversely proportional to the exciton
volume $1/L^3$.
In the case of nanotubes the electron and hole wavefunctions are
confined in two dimensions and therefore the oscillator
strength should be inversely proportional to the exciton size
$L_1$.

\begin{figure}
\includegraphics[height=2.34in,width=2.85in,angle=0]{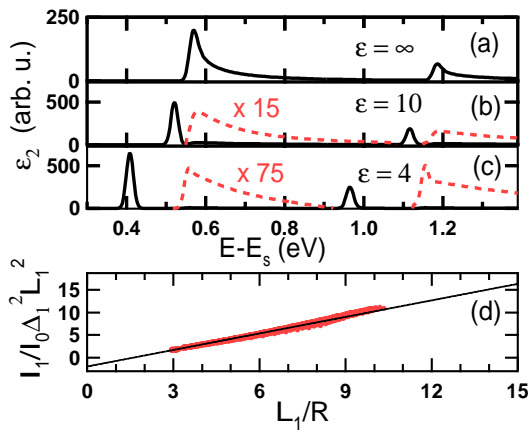}
\caption{\label{fig3} Absorption spectra $\varepsilon_2$ in (19,0)
tube in dielectric environment (a) $\varepsilon=\infty$
(equivalent to no e-h interaction), (b) $\varepsilon=10$, (c)
$\varepsilon=4$. ($E_s$ is the unknown self-energy shift.) Note
expanded scales for dotted lines in continuum region in (b) and
(c). The fractional spectral weight transfer to the first exciton
is $I_1/I_0=$0, 0.71, and 0.95 respectively. Spectra are broadened
with a Gaussian width of 0.0125 eV. (d) The scaling of spectral
weight transfer to the first exciton, $I_1$, according to
Eq.~(\protect{\ref{ii0}}), in all semiconducting tubes with
$d=1.0-2.5$ nm and $\varepsilon=2-15$ and all possible
chiralities. The best fit to Eq.~(\protect{\ref{ii0}}) (RMS
difference 3.5\%) is obtained with ${\rm A_I}=1.22$ eV$^2$nm$^2$
and ${\rm B_I}=1.61$.}
\end{figure}

Typical optical absorption
spectra are shown on Fig.~\ref{fig3}a-c, calculated for a $(19,
0)$ tube in different dielectric media. As $\varepsilon$ increases,
the spectral function converges to the non-interacting limit.
For $\varepsilon = 10$,
the spectral weight transfer to the first and
second excitons, as a fraction of the total spectral weights for
first and second bands in the non-interacting limit,
are 71\% and 55 \% respectively.
The second exciton resonance is more bound than the first,
by 63 meV vs.\ 43 meV.
The higher spectral weight transfer to the first exciton
is due not to stronger binding, but rather
to the smaller band gap $\Delta_1$.

The probability argument \cite{Elliot}
along with Eq.~(\ref{eps2}) suggest the following scaling
relation for the exciton oscillator strength:
\begin{eqnarray}
\frac{I_1}{I_0}=\frac{\rm A_I}{\Delta^2L_1R}\left(1-\frac{{\rm
B_I}R}{L_1}\right) \label{ii0}
\end{eqnarray}
where $I_0$ is the spectral weight of the first band for
non-interacting particles. The second term is a higher order
correction due to the band mixing and the non-constant band-to-band
optical density. Figure \ref{fig3}d confirms this scaling. While
$L_1$ is not directly observable, the figure is virtually
unchanged if we use the actual calculated $L_1$ values instead of
the scaling expression Eq.~(\ref{LL1}) for $L_1$. Thus $I_1 / I_0$
obeys rather well an explicit scaling relationship with
$\varepsilon/Rm$.

The radiative lifetime of the excitons in carbon nanotubes is a
key factor for possible applications in photonics and
optoelectronics. The radiative lifetime of an exciton is
inversely proportional to its oscillator strength
\cite{Stern}.
In the regime of large binding ($\varepsilon\lesssim 3$),
$I_1\approx I_0$.  Then the oscillator strength per atom
is almost independent of tube diameter and chirality,
and is equal to
$f_0\approx \left( 0.014 {\rm eV}^{-1}\right) E_{\rm exc}$.

We emphasize however that the radiative lifetime and luminescence
efficiency of nanotubes involve other factors as well. In
principle, a single exciton coherently captures spectral weight
from a macroscopic region \cite{Hanamura}; so the lifetime of an
exciton actually depends on the coherence length in the nanotube,
which in turn depends on environment and temperature.
(If the coherence length is
sufficiently large, other lengthscales such as tube length
or photon wavelength can become important.) Another important
factor is that electrons and holes are relatively unlikely to form
optically active excitons, because there are far more excitons
that are optically inactive.  These include triplet and other
dipole-forbidden excitons at {\it lower energy} than the optically
active exciton. Most importantly, only a tiny fraction of excitons
have a total momentum compatible with photon emission, so
phonon scattering plays an important role \cite{finiteQ}.

In conclusion, we have calculated optical spectra of carbon
nanotubes including the electron-hole Coulomb interaction by
solving the Bethe-Salpeter equation (\ref{eq34})
in a tight-binding wavefunction basis set. We find scaling
relations with respect to the tube radius and dielectric constant
$\varepsilon$, for the binding energy Eq.~(\ref{eq47}), exciton
size Eq.~(\ref{LL1}), and oscillator strength Eq.~(\ref{ii0}).
Thus the absorption and emission properties depend on the
dielectric media in which the nanotube is placed.
We find a
strong dependence on tube index (chirality), but only via the
effective mass.  This depends strongly on whether mod(n-m,3)=1 or
2, but is otherwise insensitive to the chiral angle.

The authors thank M. Freitag, T. Heinz, M. Hybertsen, S. G. Louie,
G. Mahan, and F. Wang for helpful discussions.


\begin{thebibliography}{}
\bibitem{dielectric}
If the exciton binding energy is much smaller the optical phonon
energy in the dielectric medium, then the static dielectric
constant $\varepsilon_0$ should be used in the scaling relations
obtained here; but if  much larger, then the screening
$\varepsilon_{\infty}$ (i.e.\ for frequencies well above the
phonon frequency) is used. A general treatment would require
explicit inclusion of exciton-phonon coupling.
\end{thebibliography}
\end{document}